\documentclass[preprint,showpacs,preprintnumbers,amsmath,amssymb]{revtex4}
\usepackage{graphics}% Include figure files
\usepackage{dcolumn}% Align table columns on decimal point
\usepackage{bm}% bold math
\begin{document}

\title{Photon spin operator and Pauli matrix}

\author{Chun-Fang Li$^{1,2}$\footnote{Email address: cfli@shu.edu.cn} and Xi Chen$^{1,3}$}

\affiliation{$^1$Department of Physics, Shanghai University, 99 Shangda Road, 200444
Shanghai, China}

\affiliation{$^2$State Key Laboratory of Transient Optics and Photonics, Xi'an Institute
of Optics and Precision Mechanics of CAS, 710119 Xi'an, China}

\affiliation{$^3$Departamento de Qu\'{\i}mica-F\'{\i}sica, UPV-EHU, Apartado 644, E-48080
Bilbao, Spain}

\date{\today}% It is always \today, today,

\begin{abstract}

Any polarization vector of a plane wave can be decomposed into a pair of mutually
orthogonal base vectors, known as a polarization basis. Regarding this decomposition as a
quasi-unitary transformation from a three-component vector to a corresponding
two-component spinor, one is led to a representation formalism for the photon spin. The
spin operator $\hat{\boldsymbol \gamma}$ defined on the space of unit spinors, referred
to as the Jones space, has only component along the wave vector and is represented by one
of the Pauli matrices in the commonly used polarization basis. It is deformed by the
quasi-unitary transformation from the spin operator that is defined on the space of unit
polarization vectors, referred to as the Pancharatnam space. On the basis of this theory,
it is shown that the Cartesian components of spin operator $\hat{\boldsymbol \gamma}$ are
mutually commutative and the spin angular momentum in units of $\hbar$ is exactly the
component of the Stokes vector along the wave vector.

\end{abstract}

%\keywords{}                       %Use showkeys class option if keyword
\maketitle

%----------------------------------------------------------------

\section{Introduction}

The spin angular momentum (SAM) of the photon is a basic quantity in the nature. Since it
was conceived~\cite{Poynting} and experimentally detected~\cite{Raman, Beth}, the photon
spin has been puzzling and still remains mysterious for the time being. First of all, the
separability of the spin from the orbital angular momentum has been a controversial issue
for a long time \cite{Akhiezer, Bere, Simmons, Jauch, Biedenharn, Cohen, van-JMO,
van-EPL, Santamato}. The key point for this is usually attributed \cite{Akhiezer, Bere,
Cohen} to the transversality of the radiation field. But there appeared in the past
decade more and more experimental evidences that the SAM is different from the orbital
angular momentum. The spin and orbital angular momentum have distinct effects on tiny
birefringent particles held in optical tweezers \cite{O'Neil-G}. The conversion of the
SAM to the orbital angular momentum was observed in anisotropic \cite{Marrucci},
isotropic \cite{Zhao} and nonlinear \cite{Mosca} media. The spin-orbit interaction of a
photon was also detected \cite{Hosten, Haefner, Qin}.

Secondly, a common opinion as for the relation between the spin and the polarization is
that the spin is the polarization itself \cite{Bere, Biedenharn, Damask, Franke,
Berry2009} or is, ambiguously, associated with the polarization \cite{Santamato, Allen,
Friese}. This seems all that we can say about the relation between the spin and the
polarization. But we are faced with a dilemma. On one hand, it was observed
\cite{van-JMO, van-EPL} that the Cartesian components of the spin operator in a second
quantized theory commutate with one another. On the other hand, the photon polarization
is usually taken as a classical analogue in quantum mechanics \cite{Sakurai} to
illustrate the non-commutability of the Cartesian components of the spin of the electron.
The question arises naturally as to how to understand the aforementioned commutativity
between the Cartesian components of the spin operator.

More importantly, though the photon spin is equal to $1$, it has only two independent
eigen components corresponding to the left-handed and right-handed circular
polarizations. So the Pauli matrices are frequently used for the spin operator in either
theoretical or experimental works, concerning, for example, the SAM transfer
\cite{Piccirillo, Choi} and the spin Hall effect \cite{Hosten, Onoda, Bliokh}. But it is
not without problem \cite{Berry1998} to regard the Pauli matrices as the Cartesian
components of the spin operator. In the first place, they do not obey the commutation
relations appropriate for a spin-$1$ particle. In the second place, they are not in
consistency with van Enk and Nienhuis' observation \cite{van-JMO, van-EPL} that the
Cartesian components of the spin operator are commutative with one another.

The purpose of this paper is to investigate whether and how the spin operator of the
photon can be expressed in terms of the Pauli matrix. As we know~\cite{Akhiezer, van-JMO,
van-EPL, Berry1998}, the spin operator that obeys the commutation relations appropriate
for a spin-$1$ particle is given by 3-by-3 matrices $\hat{\mathbf S} =\hbar \hat{\mathbf
\Sigma}$, which act on three-component vectors referred to as the photon wave function,
where $(\hat{\Sigma}_k)_{ij} =-i \epsilon_{ijk}$ with $\epsilon_{ijk}$ the
Levi-Civit\'{a} pseudotensor. It will be shown that the previously introduced 3-by-2
matrix \cite{Li2008} on the basis of transversality condition is a quasi-unitary
transformation (QUT). The spin operator $\hat{\mathbf S}$ defined on the space of
three-component vectors is transformed by this QUT into one that is defined on the space
of two-component spinors. It is the transformed spin operator that is expressible in
terms of the Pauli matrix, but in quite an unusual way. We will see that (i) the spin
operator on the spinor space has only component along the wave vector, so that its
Cartesian components commutate with one another; (ii) the spin operator on the spinor
space generates a rotation of the spinor about the wave vector; (iii) the SAM in units of
$\hbar$ is exactly the component of the Stokes vector along the wave vector.

\section{Spin operator expressed in terms of one Pauli matrix}

Consider in free space an arbitrary normalized monochromatic radiation field. The photon
wave function~\cite{Akhiezer} $\mathbf{E} (\mathbf{k})$ in momentum representation
satisfies $\int \mathbf{E}^{\dag} \mathbf{E} d\Omega =1$, where the boldfaced symbol
stands for a column vector of three components, the superscript $\dag$ denotes the
conjugate transpose, and $\mathbf{b}^{\dag} \mathbf{a}$ means the inner product
$\mathbf{b}^* \cdot \mathbf{a}$ of two complex vectors $\mathbf{a}$ and $\mathbf{b}$. Its
SAM was rigorously separated very recently \cite{Li2009} from its orbital angular
momentum on the basis of transversality condition by observing that the density of the
linear momentum can be separated in a similar way \cite{Berry2009, Bekshaev}. The
separation of the SAM from the orbital angular momentum is gauge-independent. The SAM per
photon is given by \cite{Akhiezer, Cohen, Li2009} $\mathbf{S} =-i\hbar \int \mathbf{E}^*
\times \mathbf{E} d\Omega$. By virtue of the equality \cite{Akhiezer} $\mathbf{a} \times
\mathbf{b} =i \mathbf{a}^T \hat{\mathbf{\Sigma}} \mathbf{b}$, where the superscript ``T''
denotes the transpose, it turns into
\begin{equation} \label{TSAM}
\mathbf{S} =\hbar \int \mathbf{E}^{\dag} \hat{\mathbf{\Sigma}} \mathbf{E} d\Omega,
\end{equation}
which has a quantum-mechanical interpretation as the expectation value of operator
$\hat{\mathbf{S}} =\hbar \hat{\mathbf \Sigma}$ in the state represented by the wave
function $\mathbf{E} (\mathbf{k})$. Factorizing $\mathbf{E} (\mathbf{k})$ into
$\mathbf{E} (\mathbf{k}) =\mathbf{e} (\mathbf{k}) E(\mathbf{k})$, where $\mathbf{e}
(\mathbf{k})$ is a complex unit vector determining the state of polarization and
satisfying $\mathbf{e}^{\dag} \mathbf{e} =1$, $E(\mathbf{k})$ is generally a complex
scalar function representing the amplitude and satisfying $\int |E|^2 d \Omega =1$, and
substituting into Eq. (\ref{TSAM}), one obtains $\mathbf{S} =\hbar \int \mathbf
{e}^{\dag} \hat{\mathbf{\Sigma}} \mathbf{e} |E|^2 d\Omega$. This shows that the SAM per
photon in a plane wave is \cite{Li2009}
\begin{equation}\label{spin in pw}
\mathbf{s} =\hbar \mathbf{e}^{\dag} \hat{\mathbf{\Sigma}} \mathbf{e}.
\end{equation}

\subsection{From transversality to quasi-unitary transformation}

The transversality means that the three-component polarization vector $\mathbf e$ is in
reality a two-dimensional vector \cite{Li2008}. As a result, $\mathbf e$ can be expanded
in terms of a real linearly-polarized basis as
\begin{equation}\label{QUT3}
\mathbf{e} =\alpha_u \mathbf{u} +\alpha_v \mathbf{v} \equiv m_3 \tilde{\alpha}_3,
\end{equation}
where the real unit vectors $\mathbf u$ and $\mathbf v$ are the base vectors that form a
right-handed triad with the wave vector obeying \cite{comment1}
\begin{subequations}\label{triad}
\begin{align}
  \mathbf{u}^{\dag} \mathbf{v}  & =0, \\
  \mathbf{u} \times \mathbf{v} & =\mathbf{w},
\end{align}
\end{subequations}
the unit vector $\mathbf{w} =\frac{\mathbf k}{k}$ denotes the direction of the wave
vector, and $\tilde{\alpha}_3 =\left(
                     \begin{array}{c}
                       \alpha_u \\
                       \alpha_v \\
                     \end{array}
                   \right)
$ is a complex two-component unit spinor satisfying $\tilde{\alpha}^{\dag}_3
\tilde{\alpha}_3 =1$. In Eq. (\ref{QUT3}) the base vectors $\mathbf u$ and $\mathbf v$ of
the polarization basis constitute the 3-by-2 matrix \cite{Li2008} $m_3 =\left(
                                      \begin{array}{cc}
                                        \mathbf{u} & \mathbf{v} \\
                                      \end{array}
                                    \right)
$ that satisfies
\begin{equation}\label{unitarity a}
    m^{\dag}_3 m_3 =1
\end{equation}
when Eq. (\ref{triad}) is considered. With the polarization basis $m_3$, the spinor
$\tilde{\alpha}_3$ plays the role of determining \cite{Akhiezer} the state of
polarization and will be referred to as the polarization spinor. The meaning of subscript
``3'' will be clear shortly.

The polarization basis $m_3$ is a QUT in the following sense. Firstly, it operates on a
unit spinor and yields a unit vector as Eq. (\ref{QUT3}) shows. Secondly, one readily
obtains $\tilde{\alpha}_3 =m^{\dag}_3 \mathbf{e}$ from Eq. (\ref{QUT3}) by making use of
Eq. (\ref{unitarity a}). This shows that the 2-by-3 matrix $m^{\dag}_3$ operates on a
unit vector and yields a unit spinor, indicating the following property:
\begin{equation}\label{unitarity b}
    m_3 m^{\dag}_3 =1.
\end{equation}
As a matter of fact, direct multiplication gives $m_3 m^{\dag}_3 =1- \mathbf{w}
\mathbf{w}^T$. Because the matrix $m^{\dag}_3$ always operates on the polarization vector
$\mathbf e$ that is perpendicular to $\mathbf w$, $m_3 m^{\dag}_3$ reduces to $1$. That
is to say, Eq. (\ref{unitarity b}) has taken the transversality condition into account.
$m^{\dag}_3$ is the Moore-Penrose pseudo-inverse \cite{Golub} of $m_3$. Eqs.
(\ref{unitarity a}) and (\ref{unitarity b}) express the quasi unitarity of the matrices
$m_3$ and $m^{\dag}_3$. In this regard, we will term as the Jones space the space of all
the unit polarization spinors on which $m_3$ is defined. Correspondingly, we will term as
the Pancharatnam space the space of all the unit polarization vectors on which
$m^{\dag}_3$ is defined, upon considering that Pancharatnam \cite{Pancharatnam} made the
first investigation into the physical significance of the phase of polarization vector by
exploring the interference between two non-orthogonal polarization vectors. The QUT's $m$
and $m^{\dag}$ relate these two spaces to each other.

\subsection{Spin operator defined on the Jones space}

Now we are ready to show that the spin operator on the Jones space is expressible in
terms of the Pauli matrix. Substituting Eq. (\ref{QUT3}) into Eq. (\ref{spin in pw}) and
considering Eq. (\ref{triad}), one arrives at
\begin{equation}\label{spin in m3}
    \mathbf{s} =\hbar \tilde{\alpha}^{\dag}_3 \hat{\boldsymbol \gamma}_3 \tilde{\alpha}_3,
\end{equation}
where $\hat{\boldsymbol \gamma}_3 =m^{\dag}_3 \hat{\mathbf \Sigma} m_3 =\mathbf{w}
\hat{\sigma}_3$, $\hat{\sigma}_3 =m^{\dag}_3 (\mathbf{w}^T \hat{\mathbf \Sigma}) m_3$ is
one of the three Pauli matrices \cite{Pauli matrices}, and $\mathbf{w}^T \hat{\mathbf
\Sigma}$ means the inner product of the matrix vector $\hat{\mathbf \Sigma}$ and the unit
wave vector $\mathbf w$. The spin operator $\hat{\mathbf \Sigma}$ that is defined on the
Pancharatnam space is transformed by the QUT into $\hat{\boldsymbol \gamma}_3$ that is
defined on the Jones space. This shows that $\hat{\mathbf \Sigma}$ is equivalent to
$\mathbf{w} (\mathbf{w}^T \hat{\mathbf \Sigma})$ when the QUT is taken into account. The
photon spin is thus always along $\mathbf w$. Remarkably, contrary to what might be
expected \cite{Piccirillo, Choi, Berry1998}, the spin operator $\hat{\boldsymbol \gamma}$
contains only one of the Pauli matrices, $\hat {\sigma}_3$. It describes exactly the fact
that the spin has only two eigen states of eigen values $\pm \hbar$ \cite{Mandel}. In
addition, its Cartesian components commutate with one another, in complete agreement with
van Enk and Nienhuis' conclusion \cite{van-JMO, van-EPL}.

It is the real-valuedness of the polarization basis $m_3$ that makes the spin operator on
the Jones space have the form of Pauli matrix $\hat{\sigma}_3$. This is why we adopt the
subscript ``3'' to denote that polarization basis. As we know, the change of polarization
basis is represented by a unitary transformation \cite{Mandel}. Let us first consider the
following unitary transformation,
\begin{equation*}
    U_2 =\exp \left( -i \frac{\pi}{4} \hat{\sigma}_2 \right).
\end{equation*}
Inserting the identity $U^{\dag}_2 U_2$ into Eq. (\ref{QUT3}) and letting
\begin{subequations}
\begin{align}
    m_1 & =m_3 U^{\dag}_2 =\left(
                             \begin{array}{cc}
                               \frac{\mathbf{u} +i\mathbf{v}}{\sqrt 2} & \frac{\mathbf{v} +i\mathbf{u}}{\sqrt 2} \\
                             \end{array}
                           \right), \label{PB1}\\
    \tilde{\alpha}_1 & =U_2 \tilde{\alpha}_3,
\end{align}
\end{subequations}
one has for the same polarization vector,
\begin{equation}\label{QUT1}
    \mathbf{e} =m_1 \tilde{\alpha}_1.
\end{equation}
The complex unit vectors $(\mathbf{u} +i\mathbf{v})/\sqrt{2}$ and $(\mathbf{v}
+i\mathbf{u})/\sqrt{2}$ in Eq. (\ref{PB1}) describe the two orthogonal circular
polarizations. They form the new polarization basis $m_1$, which is also a QUT.
Substituting Eq. (\ref{QUT1}) into Eq. (\ref{spin in pw}) yields
\begin{equation}\label{spin in m1}
    \mathbf{s} =\hbar \tilde{\alpha}^{\dag}_1 \hat{\boldsymbol \gamma}_1 \tilde{\alpha}_1,
\end{equation}
where $\hat{\boldsymbol \gamma}_1 =m^{\dag}_1 \hat{\mathbf \Sigma} m_1 =\mathbf{w}
\hat{\sigma}_1$, $\hat{\sigma}_1 =m^{\dag}_1 (\mathbf{w}^T \hat{\mathbf \Sigma}) m_1$,
showing that the spin operator $\hat{\mathbf \Sigma}$ on the Pancharatnam space is
transformed by $m_1$ into $\hat{\boldsymbol \gamma}_1$. As is expected, the spin operator
$\hat{\boldsymbol \gamma}_1$ on the Jones space has only component along $\mathbf w$ and
contains only one of the Pauli matrices, $\hat{\sigma}_1$.

Now we consider a second polarization-basis change represented by another unitary
transformation $U_3 =\exp \left( -i \frac{\pi}{4} \hat{\sigma}_3 \right)$. Inserting the
identity $U^{\dag}_3 U_3$ into Eq. (\ref{QUT1}) and letting
\begin{subequations}
\begin{align}
    m_2 & =m_1 U^{\dag}_3 =\left(
                             \begin{array}{cc}
                               \frac{\mathbf{v} -\mathbf{u}}{\sqrt 2} e^{i\frac{3\pi}{4}} &
                               \frac{\mathbf{u} +\mathbf{v}}{\sqrt 2} e^{i\frac{\pi}{4}} \\
                             \end{array}
                           \right), \\
    \tilde{\alpha}_2 & =U_3 \tilde{\alpha}_1,
\end{align}
\end{subequations}
one has another expression for the polarization vector,
\begin{equation}\label{QUT2}
    \mathbf{e} =m_2 \tilde{\alpha}_2.
\end{equation}
The matrix $m_2$, again a QUT, represents a third polarization basis that consists of a
pair of complex rectilinear base vectors. Upon substituting Eq. (\ref{QUT2}) into Eq.
(\ref{spin in pw}), one gets
\begin{equation*}
    \mathbf{s} =\hbar \tilde{\alpha}^{\dag}_2 \hat{\boldsymbol \gamma}_2 \tilde{\alpha}_2,
\end{equation*}
where $\hat{\boldsymbol \gamma}_2 =m^{\dag}_2 \hat{\mathbf \Sigma} m_2 =\mathbf{w}
\hat{\sigma}_2$ and $\hat{\sigma}_2 =m^{\dag}_2 (\mathbf{w}^T \hat{\mathbf \Sigma}) m_2$.
Again, the transformed spin operator $\hat{\boldsymbol \gamma}_2$ from $\hat{\mathbf
\Sigma}$ by $m_2$ is along $\mathbf w$ and contains only one of the Pauli matrices,
$\hat{\sigma}_2$.

In summary of this section we see that in a particular polarization basis $m =\left(
      \begin{array}{cc}
        \boldsymbol{\varepsilon}_1 & \boldsymbol{\varepsilon}_2 \\
      \end{array}
    \right)
$, the polarization vector is expressed as
\begin{equation}\label{QUT}
    \mathbf{e} =m \tilde{\alpha},
\end{equation}
where $\tilde{\alpha}$ is the polarization spinor associated with $m$, the base vectors
$\boldsymbol{\varepsilon}_1$ and $\boldsymbol{\varepsilon}_2$ obey
\begin{subequations}\label{base vectors}
\begin{align}
    \boldsymbol{\varepsilon}^{\dag}_1 \boldsymbol{\varepsilon}_2 & =0, \\
    \boldsymbol{\varepsilon}_1 \times \boldsymbol{\varepsilon}_2 & =\mathbf{w},
\end{align}
\end{subequations}
which guarantees that $m$ and $m^{\dag}$ satisfy
\begin{equation}\label{unitarity}
    m^{\dag} m =m m^{\dag} =1
\end{equation}
and act as the QUT's connecting the Jones and Pancharatnam spaces. The spin operator
$\hat{\mathbf \Sigma}$ on the Pancharatnam space is transformed into $\hat{\boldsymbol
\gamma} =m^{\dag} \hat{\mathbf \Sigma} m =\mathbf{w} \hat{\gamma}$ on the Jones space,
where
\begin{equation}\label{spin on JS}
    \hat{\gamma} =m^{\dag} (\mathbf{w}^T \hat{\mathbf \Sigma}) m
\end{equation}
is an Hermitian unitary matrix satisfying $\hat{\gamma}^2 =1$. This shows that the spin
operator on the Jones space is expressed by a single Hermitian unitary matrix and is
always along the direction of the wave vector. Denoting the eigen spinors of operator
$\hat{\gamma}$ by $\tilde{\alpha}_{\pm}$ satisfying $\hat{\gamma} \tilde{\alpha}_{\pm}
=\pm \tilde{\alpha}_{\pm}$ and noticing Eq. (\ref{unitarity}), one then has
$(\mathbf{w}^T \hat{\mathbf \Sigma}) \mathbf{e}_{\pm} =\pm \mathbf{e}_{\pm}$, where
$\mathbf{e}_{\pm} =m \tilde{\alpha}_{\pm}$. This means that the eigen spinors of
$\hat{\gamma}$ correspond to the eigen vectors of operator $\mathbf{w}^T \hat{\mathbf
\Sigma}$ via Eq. (\ref{QUT}). If the polarization basis is changed according to $m' =m
U^{\dag}$ and $\tilde{\alpha}' =U \tilde{\alpha}$ with a unitary transformation $U$, the
operator $\hat \gamma$ is changed as
\begin{equation}\label{transform of gamma}
    \hat{\gamma}' =U \hat{\gamma} U^{\dag}.
\end{equation}
Consequently, the Hermitian unitary matrix in the spin operator $\hat{\boldsymbol
\gamma}$ takes different forms in different polarization bases.

\section{One-to-one correspondence between the SO(3) and SU(2) rotations}

Berry \cite{Berry1998} once observed that the SAM is invariant when the polarization
vector is rotated about the wave vector. This is the case because it is just in the
direction of the wave vector. In this section we will show that a SO(3) rotation of the
polarization vector about the wave vector corresponds, via the QUT, to a SU(2) rotation
of the polarization spinor about the same axis through the same angle, a relation that is
quite different from what is known in the literature~\cite{Normand}.

Consider a polarization vector $\mathbf e$ that is given by Eq. (\ref{QUT}). A SO(3)
rotation $R(\Phi \mathbf{w}) =\exp\{-i(\mathbf{w}^T \hat {\mathbf \Sigma}) \Phi\}$ about
$\mathbf w$ through an angle $\Phi$ transforms $\mathbf e$ into
\begin{equation}\label{rotated PV}
\mathbf{e}' =R(\Phi \mathbf{w}) \mathbf{e} =R(\Phi \mathbf{w}) m \tilde{\alpha}.
\end{equation}
This can be interpreted as rotating the polarization basis about $\mathbf w$, $m' =R(\Phi
\mathbf{w}) m$, with the polarization spinor remaining unchanged. Since $\mathbf{w}^T
\hat {\mathbf \Sigma}$ commutates with $R(\Phi \mathbf{w})$, it follows from Eq.
(\ref{spin on JS}) that the spin operator on the Jones space is invariant under a
rotation of the polarization basis about $\mathbf w$,
\begin{equation*}
    \hat{\gamma}' =m'^{\dag} (\mathbf{w}^T \hat{\mathbf \Sigma}) m'
                  =m^{\dag} (\mathbf{w}^T \hat{\mathbf \Sigma}) m =\hat{\gamma}.
\end{equation*}
As a matter of fact, the two conditions in Eq. (\ref{base vectors}) do not uniquely
determine the polarization basis up to such a rotation~\cite{Mandel}. The SO(3) rotation
operator can be written as \cite{Normand}
\begin{equation*}
    R(\Phi \mathbf{w}) =\cos\Phi -i(\mathbf{w}^T \hat{\mathbf \Sigma}) \sin\Phi
    +(1-\cos\Phi) \mathbf{w} \mathbf{w}^T.
\end{equation*}
Substituting it into Eq. (\ref{rotated PV}) and noticing that the base vectors in $m$ are
perpendicular to $\mathbf w$, one has
\begin{equation*}
    \mathbf{e}'
    =\{ \cos\Phi -i(\mathbf{w}^T \hat{\mathbf \Sigma}) \sin\Phi \} m \tilde{\alpha}.
\end{equation*}
By making use of Eqs. (\ref{unitarity}) and (\ref{spin on JS}), one may rewrite it as
\begin{equation*}
    \mathbf{e}' =m \exp(-i \hat{\gamma} \Phi) \tilde{\alpha}.
\end{equation*}
This can be reinterpreted as rotating the polarization spinor about $\mathbf w$,
\begin{equation}\label{rotated PS}
    \tilde{\alpha}' =\exp(-i \hat{\gamma} \Phi) \tilde{\alpha},
\end{equation}
the generator being the spin operator, with the polarization basis remaining unchanged.
Because the generator of this SU(2) rotation is without the factor $\frac{1}{2}$ that we
encounter in the case of electrons, the rotation angle is the same as the SO(3) rotation.
This completes our proof.

The invariance of the SAM under the rotation of the polarization vector or the
polarization spinor about $\mathbf w$ means that the spin is different from the
polarization. Since the polarization state of a completely polarized plane wave can be
exactly described by the Stokes vector, we will explore in the next the relation of the
SAM with the Stokes vector.

\section{SAM is the component of the Stokes vector along $\mathbf w$}

\begin{figure}[ht]
  \includegraphics{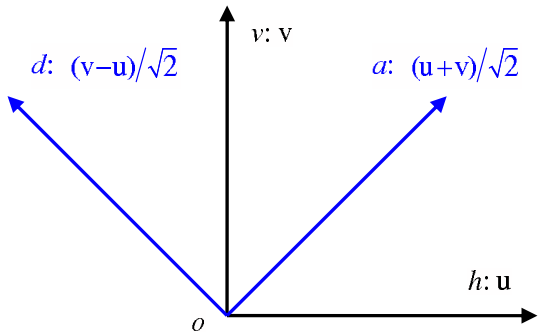}\\
  \caption{Schematic diagram illustrating the connection of the base polarization vectors
  that form the QUT $m_3$ and $m_2$ to the orientations of the linear
  polarizer for measuring the Stokes parameters $p_1$ and $p_2$.}\label{orientations}
\end{figure}
The three components of the Stokes vector $\mathbf p$ are defined by \cite{Damask}
\begin{equation}\label{defining SV}
    p_1 =\frac{I_h-I_v}{I_h+I_v}, \hspace{5pt}
    p_2 =\frac{I_a-I_d}{I_a+I_d}, \hspace{5pt}
    p_3 =\frac{I_r-I_l}{I_r+I_l},
\end{equation}
where $I_h$, $I_v$, $I_a$, and $I_d$ are the intensities of the wave measured through the
corresponding orientations of the linear polarizer as is shown in Fig.
\ref{orientations}, $I_r$ and $I_l$ are respectively the intensities of the right-handed
and left-handed circularly polarized components in the wave. The subscripts ``h'' and
``v'' mean the horizontal and vertical orientations of the polarizer, respectively, that
correspond to the base vectors $\mathbf u$ and $\mathbf v$ of the polarization basis
$m_3$. Similarly, the subscripts ``d'' and ``a'' mean that the respective orientations of
the polarizer correspond to the base vectors $\frac{\mathbf{v} -\mathbf{u}}{\sqrt 2}
e^{i\frac{3\pi}{4}}$ and $\frac{\mathbf{u} +\mathbf{v}}{\sqrt 2} e^{i\frac{\pi}{4}}$ of
the polarization basis $m_2$. With the help of polarization basis $m_3$, $p_1$ is written
in terms of the polarization spinor $\tilde{\alpha}_3$ as
\begin{equation}\label{p1}
    p_1 =\tilde{\alpha}^{\dag}_3 \hat{\sigma}_1 \tilde{\alpha}_3.
\end{equation}
In addition, the structure of the polarization basis $m_2$ allows us to express $p_2$ as
$p_2 =- \tilde{\alpha}^{\dag}_2 \hat{\sigma}_1 \tilde{\alpha}_2$. Noticing
$\tilde{\alpha}_2 =U_3 \tilde{\alpha}_1$ and $\tilde{\alpha}_1 = U_2 \tilde{\alpha}_3$,
it turns out to be
\begin{equation}\label{p2}
    p_2 =\tilde{\alpha}^{\dag}_3 \hat{\sigma}_2 \tilde{\alpha}_3.
\end{equation}
Furthermore, $p_3$ can be expressed as $p_3 =\tilde{\alpha}^{\dag}_1 \hat{\sigma}_1
\tilde{\alpha}_1$ in accordance with the polarization basis $m_1$. This is exactly the
SAM (in units of $\hbar$) as Eq. (\ref{spin in m1}) shows. A comparison between Eqs.
(\ref{spin in m1}) and (\ref{spin in m3}) leads to
\begin{equation}\label{p3}
    p_3 =\tilde{\alpha}^{\dag}_3 \hat{\sigma}_3 \tilde{\alpha}_3.
\end{equation}
Collecting Eqs. (\ref{p1})-(\ref{p3}) all together, we find that the Stokes vector can be
expressed in the polarization basis $m_3$ as
\begin{equation}\label{SV in m3}
    \mathbf{p} =\tilde{\alpha}^{\dag}_3 \hat{\boldsymbol \sigma} \tilde{\alpha}_3,
\end{equation}
in terms of the Pauli vector $\hat{\boldsymbol \sigma}$ \cite{Chru, comment2}.

In view of Eq. (\ref{SV in m3}), the Pauli vector should be regarded as the polarization
operator. This is why the polarization of photons can be compared \cite{Sakurai} to the
spin of electrons. Since the Stokes vector is invariant under the change of polarization
basis, the detailed form of the Pauli vector is dependent on the choice of polarization
basis. If the polarization basis is changed according to $\tilde{\alpha}' =U
\tilde{\alpha}_3$ with $U$ a unitary transformation, the Stokes vector appears to be
$\mathbf{p} =\tilde{\alpha}'^{\dag} U \hat{\boldsymbol \sigma} U^{\dag} \tilde{\alpha}'$.
Letting be $\hat{\boldsymbol \sigma}'$ the Pauli vector in the new polarization basis,
\begin{equation}\label{change of PV}
    \hat{\boldsymbol \sigma}' =U \hat{\boldsymbol \sigma} U^{\dag},
\end{equation}
one has $\mathbf{p} =\tilde{\alpha}'^{\dag} \hat{\boldsymbol \sigma}' \tilde{\alpha}'$.
In a word, denoting respectively by $\hat{\boldsymbol \sigma}$ and $\tilde \alpha$ the
Pauli vector and the polarization spinor in a particular polarization basis, the Stokes
vector is given by
\begin{equation}\label{Stokes vector}
    \mathbf{p} =\tilde{\alpha}^{\dag} \hat{\boldsymbol \sigma} \tilde{\alpha}.
\end{equation}

We have shown that the third component of the Stokes vector is equal to the SAM. Now that
the SAM is in the direction of the wave vector, it is exactly the component of the Stokes
vector along the wave vector. This is easily proven. Suppose that the spin operator on
the Jones space is given by
\begin{equation}\label{gamma from sigma}
    \hat{\gamma} =\mathbf{w}^T \hat{\boldsymbol \sigma}
\end{equation}
with $\hat{\boldsymbol \sigma}$ the Pauli vector in a particular polarization basis. Eq.
(\ref{gamma from sigma}) guarantees that the component of the Stokes vector along
$\mathbf w$, $p_3 =\tilde{\alpha}^{\dag} (\mathbf{w}^T \hat{\boldsymbol \sigma})
\tilde{\alpha}$, is invariant under the rotation of the polarization spinor about
$\mathbf w$ by virtue of Eq. (\ref{rotated PS}), the same as the SAM is. So the component
of the Stokes vector along $\mathbf w$ is the SAM. Jauch and Rohrlich~\cite{Jauch} once
found that the SAM is equal to one component of the Stokes vector in a second
quantization theory of the radiation field. Unfortunately, their result received little
attention.

\section{Concluding remarks}

In conclusion, we have shown that the photon spin operator on the Jones space is given by
$\hat{\boldsymbol \gamma} =\mathbf{w} \hat{\gamma}$ in terms of an Hermitian unitary
matrix $\hat{\gamma}$. It is in the direction of the wave vector. Its Cartesian
components are commutative with one another. This operator is obtained through
transforming the spin operator $\hat{\mathbf \Sigma}$ that is defined on the Pancharatnam
space and satisfies the appropriate commutation relations by making use of a QUT that is
associated with a particular polarization basis. The form of matrix $\hat{\gamma}$
depends on the choice of the polarization basis. In the commonly used polarization bases
denoted by $m_1$, $m_2$, and $m_3$, $\hat{\gamma}$ takes the form of the Pauli matrices
$\hat{\sigma}_1$, $\hat{\sigma}_2$, and $\hat{\sigma}_3$, respectively. On the basis of
this theory, we found that a SO(3) rotation of the polarization vector about the wave
vector corresponds to a SU(2) rotation of the polarization spinor through the same angle
about the same axis. It is very interesting to note that the generator of the former
rotation is the component of spin operator $\hat{\mathbf \Sigma}$ along the wave vector
and that of the latter rotation is the spin operator $\hat{\boldsymbol \gamma}$ itself.
Furthermore, the SAM is just the component of the Stokes vector along the wave vector.

The theory advanced here for the photon spin on the Jones space is a probabilistic one
that is compatible with the quantum mechanical description \cite{Akhiezer} of a photon
state. Upon transforming from the Pancharatnam space to the Jones space by the QUT, one
arrives at the two-component polarization spinor, which is much suitable to deal with the
angular momentum problem of the radiation field. It is noted that the two conditions in
Eq. (\ref{triad}) or Eq. (\ref{base vectors}) leave the polarization basis undetermined
up to a rotation about the wave vector. It is the degree of freedom of this rotation in
defining the QUT~\cite{Li2008} that makes it necessary for us to introduce a fixed unit
vector to represent a vector electromagnetic beams. That vector plays a very important
role in describing the orbital angular momentum of light beams~\cite{Li2009} and in
understanding the spin Hall effect of light~\cite{Li2009-2}. A general formalism for the
spin and orbital angular momentum on the Jones space is under preparation.

\section*{Acknowledgments}

CFL would like to thank Masud Mansuripur of the University of Arizona for his helpful
discussions. This work was supported in part by the National Natural Science Foundation
of China (60877055 and 60806041), the Shanghai Educational Development Foundation
(2007CG52), and the Shanghai Leading Academic Discipline Project (S30105).


\begin{thebibliography}{99}

\bibitem{Poynting} J. H. Poynting, Proc. R. Soc. Lond. A 82, 560 (1909).
\bibitem{Raman} C.V. Raman and S. Bhagavantam, Nature 129, 22 (1932).
\bibitem{Beth} R.A. Beth, Phys. Rev. 50, 115 (1936).
\bibitem{Akhiezer} A. I. Akhiezer and V. B. Berestetskii, Quantum electrodynamics
(Interscience Publishers, New York, 1965).
\bibitem{Bere} V. B. Berestetskii, E. M. Lifshitz, and L. P. Pitaevskii, Quantum
electrodynamics, 2nd ed. (Pergamon Press Ltd., NY, 1982).
\bibitem{Simmons} J. W. Simmons and M. J. Guttmann, States, Waves, and Photons
(Addison-Wesley, Reading, MA, 1970).
\bibitem{Jauch} J. M. Jauch and F. Rohlich, The Theory of Photons and Electrons
(Springer-Verlag, Berlin, 1976).
\bibitem{Biedenharn} L. C. Biedenharn and J. D. Louck, Angular Momentum in Quantum Physics:
Theory and Application (Addison-Wesley, Massachusetts, 1980).
\bibitem{Cohen} C. Cohen-Tannoudji, J. Dupont-Roc, and G. Grynberg, Photons and Atoms
(Wiley, New York, 1989).
\bibitem{van-JMO} S. J. van Enk and G. Nienhuis, J. Mod. Opt. 41, 963 (1994).
\bibitem{van-EPL} S. J. van Enk and G. Nienhuis, Europhys. Lett. 25, 497 (1994).
\bibitem{Santamato} E. Santamato, Fortschr. Phys. 52, 1141 (2004).
\bibitem{O'Neil-G} A. T. O'Neil, I. MacVicar, L. Allen, and M. J. Padgett, Phys. Rev. Lett.
88, 053601 (2002); V. Garc\'{e}s-Ch\'{a}vez, D. McGloin, M. J. Padgett, W. Dultz, H.
Schmitzer, and K. Dholakia, Phys. Rev. Lett. 91, 093602 (2003).
\bibitem{Marrucci} L. Marrucci, C. Manzo, and D. Paparo, Phys. Rev. Lett. 96, 163905 (2006).
\bibitem{Zhao} Y. Zhao, J. S. Edgar, G. D. M. Jeffries, D. McGloin, and D. T. Chiu,
Phys. Rev. Lett. 99, 073901 (2007).
\bibitem{Mosca} S. Mosca et. al., Phys. Rev. A 82, 043806 (2010).
\bibitem{Hosten} O. Hosten and P. Kwiat, Science 319, 787 (2008).
\bibitem{Haefner} D. Haefner, S. Sukhov, and A. Dogariu, Phys. Rev. Lett. 102, 123903 (2009).
\bibitem{Qin} Y. Qin, Y. Li, H. He, and Q. Gong, Opt. Lett. 34, 2551 (2009).

\bibitem{Damask} J. N. Damask, Polarization optics in telecommunications (Springer
Science+Business Media, Inc., New York, 2005).
\bibitem{Franke} S. Franke-Arnold, L. Allen, and M. Padgett, Laser Photonics Rev. 2, 299 (2008).

\bibitem{Berry2009} M. V. Berry, J. Opt. A: Pure Appl. Opt. 11, 094001 (2009).

\bibitem{Allen} L. Allen, M. W. Beijersbergen, R. J. C. Spreeuw, and J. P. Woerdman,
Phys. Rev. A 45, 8185 (1992).
\bibitem{Friese} M. E. J. Friese, J. Enger, H. Rubinsztein-Dunlop, and N. R. Heckenberg,
Phys. Rev. A 54, 1593 (1996).
\bibitem{Sakurai} J. J. Sakurai, Modern Quantum Mechanics (Addison-Wesley, New York, 1985).
\bibitem{Piccirillo} B. Piccirillo, C. Toscano, F. Vetrano, and E. Santamato, Phys. Rev.
Lett. 86, 2285 (2001).
\bibitem{Choi} Hyunhee Choi, J. H. Woo, and J. W. Wu, J. Opt. Soc. Am. B 25, 491 (2008).
\bibitem{Onoda} M. Onoda, S. Murakami, and N. Nagaosa, Phys. Rev. Lett. 93, 083901 (2004).
\bibitem{Bliokh} K. Y. Bliokh and Y. P. Bliokh, Phys. Rev. Lett. 96, 073903 (2006);
Phys. Rev. E 75, 066609 (2007).
\bibitem{Berry1998} M. V. Berry, Proc. SPIE 3487, 6 (1998).
\bibitem{Li2008} C.-F. Li, Phys. Rev. A 78, 063831 (2008).

\bibitem{Li2009} C.-F. Li, Phys. Rev. A 80, 063814 (2009).
\bibitem{Bekshaev} A. Y. Bekshaev and M. S. Soskin, Opt. Commun. 271, 332 (2007).
\bibitem{comment1} For later convenience, the transpose of a real base vector is written as
its conjugate transpose, $\mathbf{u}^T =\mathbf{u}^{\dag}$.
\bibitem{Golub} G. H. Golub and C. F. Van Loan, Matrix computations, 3rd ed.
(Johns Hopkins, Baltimore, 1996).
\bibitem{Pancharatnam} S. Pancharatnam, Proc. Ind. Acad. Sci. A 44, 247 (1956).
\bibitem{Pauli matrices} In this paper, the Pauli matrices are denoted as $\hat{\sigma}_1 =\left(
                                                                             \begin{array}{cc}
                                                                               1 & 0  \\
                                                                               0 & -1 \\
                                                                             \end{array}
                                                                           \right)
$, $\hat{\sigma}_2 =\left(
                      \begin{array}{cc}
                        0 & 1 \\
                        1 & 0 \\
                      \end{array}
                    \right)
$, and $\hat{\sigma}_3 =\left(
                          \begin{array}{cc}
                            0 & -i \\
                            i & 0 \\
                          \end{array}
                        \right)
$.
\bibitem{Mandel} L. Mandel and E. Wolf, Optical coherence and quantum optics (Cambridge
University Press, New York, 1995).
\bibitem{Normand} J.-M. Normand, A Lie group: rotations in quantum mechanics
(North-Holland Publishing Company, Amsterdam, 1980).
\bibitem{Chru} D. Chru\'{s}ci\'{n}ski and A. Jamio{\l}kowski, Geometric Phases in
Classical and Quantum Mechanics (Birkh\"{a}user Boston, New York, 2004).
\bibitem{comment2} This fact was also pointed out in Ref. \cite{Damask}. Unfortunately,
the author misinterpreted the Pauli vector as spin vector.
\bibitem{Li2009-2} C.-F. Li, Phys. Rev. A 79, 053819 (2009).

\end{thebibliography}
\end{document}